\RequirePackage{fix-cm}            
\documentclass{ieee-ojvt}
\usepackage{cite}
\usepackage{amsmath,amssymb,amsfonts}
\usepackage{bm}
\usepackage{dsfont}
\usepackage{bbm}
\let\labelindent\relax
\usepackage{enumitem}
\usepackage{graphicx}
\usepackage{xcolor}
\usepackage{array}
\usepackage{tabularx}
\usepackage{float}
\usepackage{textcomp}
\renewcommand{\arraystretch}{1.2}
\usepackage{algorithm}
\usepackage{algpseudocode}
\usepackage{comment}
\usepackage{etoolbox}
\usepackage{setspace}
\usepackage{hyperref}

\def\BibTeX{{\rm B\kern-.05em{\sc i\kern-.025em b}\kern-.08em
    T\kern-.1667em\lower.7ex\hbox{E}\kern-.125emX}}
\AtBeginDocument{\definecolor{ojcolor}{cmyk}{0.93,0.59,0.15,0.02}}
\def\OJlogo{\vspace{-12pt}\includegraphics[height=24pt]{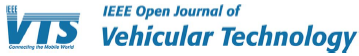}}
\usepackage{titlesec}
\usepackage{xcolor}
\titleformat{\section}{\normalfont\sffamily\bfseries\color{ojcolor}}{\thesection}{1em}{}%
\titlespacing{\subsubsection}{0pt}{1.0ex plus 0.2ex minus 0.2ex}{0.3ex}
\begin{document}
\receiveddate{XX Month, XXXX}
\reviseddate{XX Month, XXXX}
\accepteddate{XX Month, XXXX}
\publisheddate{XX Month, XXXX}
\currentdate{16 November 2026}
\doiinfo{OJVT.2026.0627000}

\title{DNN-Assisted Deterministic Policy Learning for Adaptive IRS-NOMA-OWC Networks}

\author{Ahrar N. Hamad \authorrefmark{1} (Member, IEEE), Ahmad A. Qidan \authorrefmark{1} (Senior Member, IEEE), Taisir E. H. El-Gorashi \authorrefmark{1} (Senior Member, IEEE), and Jaafar M. H. Elmirghani \authorrefmark{1} (Fellow, IEEE)}
\affil{Department of Engineering, King’s College London, London, WC2R 2LS, UK}
\corresp{CORRESPONDING AUTHOR: Ahrar N. Hamad (e-mail: ahrar.hamad@kcl.ac.uk).}
\authornote{This work has been supported  by the Engineering and Physical Science Research Council (EPSRC),  by the INTERNET project
under Grant EP/H040536/1,  by the STAR project under Grant
EP/K016873/1,  by the TOWS project under Grant EP/S016570/1, and by the TITAN project under Grant EP/X04047X/2. All data are provided in full in the results section of this article.
  All authors are with the Department of Engineering, King’s College London, London, WC2R 2LS, UK.}
\markboth{DNN-Assisted Deterministic Policy Learning for Adaptive IRS-NOMA-OWC Networks}{\textit{Ahrar et al.}}

\begin{abstract}
Intelligent reflecting surface (IRS) has been considered a promising technology in optical wireless communication (OWC) networks due to its potential to overcome the line-of-sight (LoS) blockage issue and enhance coverage. In this paper, a dynamic IRS-assisted indoor non-orthogonal multiple access (NOMA)-OWC network is modeled, and an optimization problem is formulated to maximize the sum rate subject to the quality-of-service (QoS) requirements of multiple mobile users, by jointly optimizing the power allocation and the IRS mirror orientations, from which the user's decoding order is induced rather than being fixed. Since the problem is non-convex and the channel conditions vary with user mobility, conventional optimization methods may incur considerable computational overhead. Therefore, a two-stage deep neural network-deep reinforcement learning (DNN-DRL) framework is proposed. In the first stage, the DNN estimates the user's positions and orientations in real time, removing the need for perfect knowledge of the user state. In the second stage, a deterministic policy gradient (DPG)-based DRL algorithm uses these estimates to jointly optimize the power allocation and IRS mirror orientations, with the decoding order updated according to the resulting effective channel gains. The results show that the proposed framework approaches the performance upper bounds and outperforms standard DRL algorithms, conventional optimization methods, and deployments without IRS in terms of sum rate and fairness, while achieving faster convergence. Furthermore, online adaptation of the decoding order maintains higher QoS satisfaction than a fixed decoding order, particularly under high user mobility. 
\end{abstract}

\begin{IEEEkeywords}
 Optical wireless communication (OWC), intelligent reflecting surface (IRS), non-orthogonal multiple access (NOMA), reinforcement learning (RL).\end{IEEEkeywords}
\maketitle

\setlength{\textfloatsep}{8pt}
\section{Introduction} \label{sec:6.1}
\IEEEPARstart{T}{he} exponential growth of connected devices in the sixth-generation (6G) networks presents unprecedented challenges, with global networks projected to support over 500 billion devices by 2030 \cite{yastrebova2018future}, creating challenges in existing infrastructure and spectrum resources. Recently, optical wireless communication (OWC) has emerged as a promising complement to traditional radio frequency (RF) networks, supporting a massive number of connections while enabling multi-gigabit data rates \cite{Qidan}. In particular, visible light communication (VLC) has attracted significant research attention for indoor wireless data transmission, as it uses the existing lighting infrastructure to deliver high-speed data transmission alongside illumination \cite{krishnamoorthy2025optical}. However,  the performance of VLC networks depends critically on maintaining a direct line-of-sight (LoS) link between transmitters and receivers, which is difficult to guarantee in practical indoor environments where shadowing and physical obstructions cause severe signal degradation.

To address this limitation, intelligent reflecting surface (IRS) technology has emerged as a promising solution for VLC systems operating under LoS blockages and user mobility~\cite{10295484}. An IRS consists of passive reflective elements whose orientations can be adjusted to steer incident signals towards mobile users, thereby extending AP coverage and improving achievable data rates while offering energy-efficiency advantages over traditional active relays~\cite{sun2022joint}. However, optimising the reflected signal requires knowledge of each user's position and device orientation, since the reflected channel gain depends on the mirror orientation, the incidence angle at the IRS, and the orientation of the receiving device. Maintaining this information is challenging in dynamic indoor environments, where users move continuously and their devices may change orientation. Consequently, the required IRS configuration must be updated to account for variations in user positions and orientations.

A further challenge in VLC networks arises from the limited modulation bandwidth of light-emitting diodes (LEDs), which restricts the achievable data rates~\cite{palitharathna2024handover}. To improve spectral efficiency, non-orthogonal multiple access (NOMA) enables multiple users to share the same frequency resources. Moreover, NOMA provides flexible transmit power allocation, allowing users' achievable rates to be tailored to heterogeneous quality-of-service (QoS) requirements. However, exploiting this flexibility requires jointly determining the power allocation and decoding order, as both depend on the users' effective channel gains and directly influence the achievable rates. In conventional NOMA-based VLC systems, the decoding order is typically determined according to the users' instantaneous channel gains. However, this approach does not explicitly account for heterogeneous QoS requirements, where certain users may require higher data rates regardless of their channel conditions. Consequently, a decoding order based solely on channel gains may be suboptimal for satisfying these requirements. To address this limitation, recent research has investigated the integration of IRS into indoor NOMA-VLC systems to enhance the effective channel conditions of users experiencing poor signal quality~\cite{11059850}. Since the decoding order depends on the effective channel gains, reshaping these gains through the IRS provides an opportunity to influence the decoding order and adapt it to the users' instantaneous channel conditions and QoS requirements.

The resulting optimization problem in IRS-assisted NOMA-OWC is non-convex and involves coupled continuous decision variables that must be optimized under time-varying channel conditions. Conventional iterative algorithms typically rely on accurate channel information and may converge to local optima, while their computational complexity can limit their suitability for dynamic mobile environments. Consequently, adaptive learning-based approaches are needed to jointly optimize NOMA power allocation and IRS mirror orientations, while accounting for the resulting decoding order. Machine learning (ML) offers a data-driven alternative by learning mappings and policies from measurements and interactions with the environment, reducing the reliance on explicit analytical models and computationally intensive online optimization. For user-state estimation, deep neural networks (DNNs) can learn complex nonlinear mappings between physical-layer measurements and the underlying user positions and orientations~\cite{arfaoui2021invoking}. For resource allocation and control, deep reinforcement learning (DRL) agents can learn adaptive decision-making policies through reward-driven optimization and interaction with the environment. This makes DRL suitable for problems involving time-varying channel conditions and continuous, high-dimensional decision spaces. Once trained, DRL agents can generate actions with relatively low online computational overhead, enabling rapid adaptation of IRS mirror orientations and NOMA power allocation. Furthermore, learning-based policies can account for variations caused by user mobility and blockage, provided that these conditions are adequately represented during training.

\vspace{-2mm}
\subsection{Related Works and Motivation}
Resource allocation in NOMA-based OWCs has been extensively studied, since transmit power allocation is a key mechanism for serving multiple users simultaneously. In~\cite{11673253}, a dynamic power allocation scheme was proposed for NOMA-based transmission in 6G OWC networks, where transmit power is allocated according to users' channel gains to satisfy their rate requirements. However, the achievable rates remain constrained by the quality of the available propagation paths, as power allocation alone cannot fully compensate for severe channel degradation caused by LoS blockage. This limitation has motivated the integration of IRSs into NOMA-VLC systems, where the reflecting elements provide additional propagation paths, improving the effective channel gains of users experiencing poor signal quality~\cite{liu2023sum, maraqa2023optical,10892193}.
In~\cite{liu2023sum}, the achievable sum rate of an optical IRS-aided NOMA-VLC system was maximized by optimizing the reflection matrix, where the IRS was described by an association matrix and the resulting binary programming problem was solved by an iterative algorithm. In~\cite{maraqa2023optical}, the concept of an optical simultaneous transmission and reflection reconfigurable intelligent surface (OSTAR-RIS) was introduced to overcome LoS blockage in a VLC system using power-domain NOMA and rate-splitting multiple access, where a sine-cosine algorithm was proposed to solve a sum rate maximization problem. In~\cite{10892193}, the resource allocation problem for energy efficiency maximization was formulated for an optical IRS-assisted NOMA-VLC system in which the IRS configuration and the power allocation were optimized jointly, and the problem was decomposed into sub-problems solved iteratively using a three-relaxed iterative approach. In these works, however, the IRS is optimized for static user configurations, requiring the optimization to be repeated whenever the user positions change. This limits their suitability for dynamic environments with continuous user mobility.

Other studies have treated the decoding order as an explicit optimization variable. For example, in~\cite{liu2024optical}, passive and active beamforming were jointly optimized in an RIS-assisted NOMA MISO VLC system to maximize the sum rate subject to decoding-order constraints. In~\cite{11100494}, the IRS configuration was optimized for a NOMA-based MISO VLC system using binary assignment matrices that allocate each reflecting unit to a transmitter-user pair, while power allocation follows a uniform rule and the decoding order is imposed in advance through rate constraints. In~\cite{abumarshoud2022intelligent}, the IRS reflection coefficients, decoding order, and power allocation were jointly optimized to minimize the bit error rate using an adaptive-restart genetic algorithm. A common limitation of these approaches is that the decoding order is explicitly treated as a discrete optimization variable and solved using conventional optimization algorithms, increasing the computational complexity and limiting their suitability for real-time operation in dynamic environments.

Taken together, the aforementioned studies addressed power allocation, IRS configuration, and decoding order either individually or in pairs, assuming a static problem instance. In contrast, jointly optimizing these tightly coupled variables under real-time channel variations requires adaptive methods capable of capturing the dynamics of the wireless environment and responding within the channel coherence time. In this context, deep learning offers a means of obtaining the user state without the signal overhead of explicit channel estimation. In~\cite{arfaoui2021invoking}, a DNN was trained to jointly estimate the position and orientation of a LiFi user from the received signal strengths measured across multiple APs. For resource allocation, DRL has been widely adopted, as it learns a control policy through interaction with the environment and therefore adapts to channel variations without requiring the problem to be re-solved at each instance. In~\cite{Hammadi2022DQN}, a deep Q-network (DQN) algorithm was deployed to optimize an indoor NOMA-VLC network through joint power allocation and LED transmission angle tuning. The DQN algorithm was also proposed in~\cite{ciftler2021dqn} to optimize the power allocation in hybrid RF/VLC networks. Value-based methods such as DQN, however, are designed for discrete action spaces and cannot be applied to the continuous mirror orientations and power levels without discretization, which introduces approximation errors and results in a number of discrete actions that grows exponentially with the number of users and mirrors.

To overcome this limitation, the deep deterministic policy gradient (DDPG) algorithm employs an actor–critic architecture that is well suited to continuous action spaces, making it suitable for resource allocation in dynamic multi-user NOMA-VLC environments~\cite{lillicrap2015continuous}. For example, a distributed DDPG algorithm was developed in~\cite{9667214} for hybrid RF/VLC networks to optimize downlink transmit power allocation. In~\cite{9784887}, DDPG was used to jointly optimize the beamforming weights and mirror orientations in an IRS-assisted VLC system to maximize the secrecy capacity. In our prior work~\cite{ahrar}, a two-agent DRL algorithm was proposed to improve the energy efficiency and fairness of an indoor IRS-assisted NOMA-VLC system. However, these studies determine the decoding order based on the instantaneous channel gains and assume perfect knowledge of the users' positions and orientations. Such assumptions are restrictive in practical deployments, where user mobility and device orientation changes lead to time-varying channel conditions, while user states cannot be known exactly. Furthermore, the decoding order is not explicitly adapted by the DRL agent but is determined by the resulting channel gains. Consequently, the joint impact of user-state uncertainty, power allocation, IRS orientations, and the resulting decoding order has not been adequately addressed. This motivates the proposed two-stage learning framework, which combines DNN-based user-state estimation with DDPG-based resource allocation and IRS control for adaptive IRS-assisted NOMA-VLC networks. 

\vspace{-3mm}
\subsection{Contribution}
Motivated by the aforementioned challenges, we model a dynamic indoor IRS-assisted NOMA-OWC system that takes into account multi-user mobility, LoS blockage, and heterogeneous traffic demands. To the best of our knowledge, this is the first work in which the NOMA decoding order is induced online from the IRS-reconfigured channel gains, jointly with the power allocation and the mirror orientation, rather than being fixed by the instantaneous gains or optimized offline as a discrete variable. To solve the resulting non-convex problem, a two-stage DNN-DRL framework is proposed, in which the user state estimation is separated from the resource control. The main contributions of this work are listed as follows:

\begin{itemize}
\item An optimization problem is formulated to maximize the sum rate while guaranteeing the minimum QoS requirements of each user in a dynamic multi-user IRS-assisted NOMA-OWC system under user mobility, time-varying channel conditions, and LoS blockage. The joint optimization of the power allocation and IRS mirror orientations is non-convex, with a combinatorial structure arising from the decoding order, which is induced by the IRS-reconfigured channel gains rather than optimized as an independent variable.

\item A  novel two-stage DNN-DRL framework is developed to solve the formulated problem in real-time. In the first stage, a DNN estimates the users' positions and orientations from the received signal strengths over both the LoS and the IRS-reflected links, which improves the estimation accuracy under blockage. In the second stage, a DPG-based DRL algorithm uses these estimates to jointly optimize the power allocation and the IRS mirror orientations, which reshape the effective channel gains from which the decoding order is induced.

\item The problem is reformulated as a Markov decision process (MDP), and the resulting framework is evaluated through extensive simulations in an indoor scenario with multiple users having heterogeneous QoS requirements and moving at varying velocities following the random-waypoint (RWP) mobility model. The results show that the proposed framework outperforms conventional convex optimization, standard DRL baselines, and deployments without IRS or under orthogonal access in terms of convergence, achievable sum rate, fairness, and QoS satisfaction.
\end{itemize}
\vspace{-1mm}
The rest of this paper is organized as follows: Section \ref{sec:sm} presents the system model for the indoor downlink IRS-aided NOMA-VLC network. In Section \ref{6.3}, the optimization problem formulation for joint power allocation, IRS mirror orientation and decoding order is formulated to maximize the sum rate. Section \ref{6.4} introduces our proposed intelligent framework for solving the optimization problem. In Section \ref{sec:6.5}, simulation results are presented. Finally, conclusions are provided in Section \ref{Sec:6.7}.

\vspace{-3mm}
\section{System Model} \label{sec:sm}
\begin{figure}[t]
    \centerline{\includegraphics[width=2.5in]{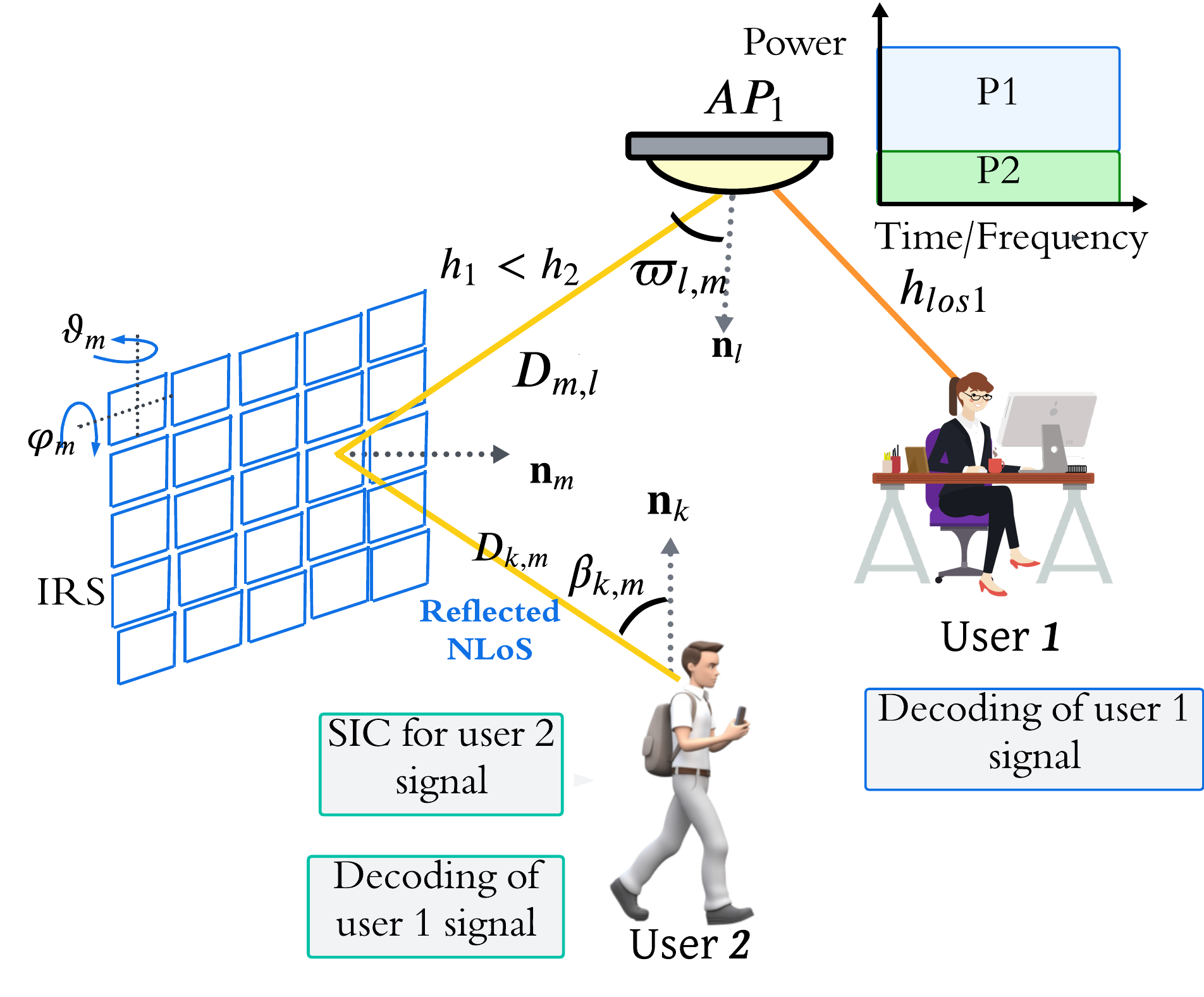}}
    \caption{IRS-aided downlink NOMA-OWC system model.}
    \label{fig6:system}
    \vspace{-1mm}
\end{figure}

We consider a downlink IRS-assisted NOMA-OWC system in an indoor environment of dimensions $L_r \times W_r \times H_r$, as shown in Fig.~\ref{fig6:system}. The system consists of a set of LED APs, ${L} = \{1, \ldots, \mathcal{L}\}$, mounted vertically downward on the ceiling and each equipped with a photodetector, serving a set of $K$ distributed mobile users, $K = \{1, \ldots, \mathcal{K}\}$. An IRS mirror array is mounted on one of the walls, consisting of $M$ passive reflecting elements, $M = \{1, \ldots,\mathcal{M}\}$. Each mirror $m$ has a rectangular area $dA_m = h_m \times w_m$, where $h_m$ and $w_m$ denote its height and width, respectively, and a spacing $u_m$ is maintained between adjacent mirrors~\cite{sun2022joint}. Each mirror can be adjusted via micro-electro-mechanical systems (MEMS) technology~\cite{10080950} along two rotational angles, the roll angle $\varphi_m$ and the yaw angle $\vartheta_m$, enabling it to steer the signal reflected from the AP toward the users. On the communication floor, the mobile users have dynamic traffic demands and move at variable velocities within $[v_{\min}, v_{\max}]$ following the RWP mobility model~\cite{bettstetter2004stochastic}. Each user’s device has a random orientation that can point in any direction. The system operates in a bidirectional mode, with a VLC downlink for data transmission and an infrared (IR) uplink used for position and orientation estimation. 
\vspace*{-2mm}
\subsection{VLC channel model}\label{AA}
\textit{1) LoS channel gain:} The channel gain of the direct LoS between user $k$ and LED $l$ can be given by
\begingroup
{\small
\begin{equation}\label{eq:losch}
H_{k,l}^{\text{LoS}} = 
\begin{cases} 
\frac{(n+1) A_r  g(\psi_c) T_s \cos^n(\varpi_{k,l}) \cos(\delta_{k,l})}{2 \pi D_{k,l}^2}, & 0 \leq \delta_{k,l} \leq \psi_c \\
 \makebox[4.4cm][c]{0} , & \delta_{k,l} > \psi_c
\end{cases}
\end{equation}}
\endgroup

\noindent where \( n = -\ln(2)/\ln(\cos(\phi_{1/2})) \) is the order of Lambertian emission, \(\phi_{1/2}\) is the LED half-intensity radiation angle. \( A_r \) is the physical area of the detector, $g(\psi_c)$ is the optical concentrator gain,  $T_s$ is the optical filter transmittance, \( \varpi_{k,l} \) is the radiation angle between the normal to the AP \( l \) and the irradiance ray of the user \( k \), \( \delta_{k,l} \) is the angle between the normal of the photodetector and the incident ray, and \( D_{k,l} \) is the distance between AP \( l \) and user \( k \). Note that, the incidence angle \( \delta_{k,l} \) must be within a range from 0 to the acceptance semi-angle of the concentrator \( \psi_c \), to guarantee that the LoS signal is detected by the receiver; otherwise, no signal is received, i.e., $H_{k,l}^{\text{LoS}}$ = 0. The impact of random device orientation on the signal propagation can be expressed using the cosine of $\delta_{k,l}$ given by 
\vspace{-1mm}

{\small
\begin{equation}
\label{LoSo}
\begin{aligned}
    \cos(\delta_{k,l}) = & \bigg(\frac{x_l - x_k}{D_{k,l}}\bigg) \cos(\varsigma) \sin(\Theta)  
    + \bigg(\frac{y_l - y_k}{D_{k,l}}\bigg)\\ 
    &\sin(\varsigma) \sin(\Theta) + \bigg(\frac{z_l - z_k}{D_{k,l}}\bigg) \cos(\Theta),
\end{aligned}
\end{equation}}

\noindent where $(x_l, y_l, z_l)$ and $(x_k, y_k, z_k)$ are the coordinates of LED $l$ and user $k$, respectively. The angles $\varsigma$ and  $\Theta$ are the azimuth and polar angles of the receiver.

\textit{2) IRS NLoS channel gain:}
The channel gain of the indirect communication signal is reflected by the mirrors of the IRS. The received channel gain by user \( k \in K\) from  mirror \( m \in M\) reflecting the signal transmitted from  AP \( l \) is expressed as
{\small
\begin{equation} \label{IRSCH}
H_{k,m,l}^{\text{IRS}} =
\begin{cases}
\dfrac{(n+1) \rho_m A_r \, dA_m \, g(\psi_c) T_s}{2\pi^2 D_{m,l}^{2} D_{k,m}^{2}}
\cos^n(\varpi_{m,l}) \cos(\beta_{m,l}) \\
\qquad \times \cos(\varpi_{k,m}) \cos(\beta_{k,m}), \
\qquad\qquad 0 \leq \beta_{k,m} \leq \psi_c\\ 
\makebox[3.7cm][c]{0} , \qquad\qquad \text{otherwise.}
\end{cases}
\end{equation}}

\noindent where \( \rho_m \) is the mirror reflection coefficient, \( dA_m \) is the area of mirror \( m \), \( D_{m,l} \) is the distance between AP \( l \) and mirror \( m \), and \( D_{k,m} \) is the distance between mirror \( m \) and user \( k \). The $\varpi_{m,l}$ is the radiation angle at AP $l$ toward mirror $m$, and $\beta_{m,l}$ is the incidence angle at mirror $m$,  $\varpi_{k,m}$ is the irradiance angle at mirror $m$ toward user $k$. 
$\beta_{k,m}$ is the incidence angle of the signal reflected from mirror \( m \) to user \( k \). In this work, the orientation of each mirror $m$ is controlled by the roll $\varphi_m$ and yaw $\vartheta_m$ angles, which define its unit normal vector given by
\vspace{-1mm}
{\small
\begin{equation}
\mathbf{n}_{m}=
\big[\sin(\varphi_{m})\cos(\vartheta_{m}),\
\cos(\varphi_{m})\cos(\vartheta_{m}),\
\sin(\vartheta_{m})\big]^{T}.
\label{eq:mirror_normal}
\end{equation}}

Since the LEDs are mounted vertically downward on the ceiling, their normal is
$\mathbf{n}_{l}=[0,0,-1]^{T}$ and the radiation angle $\cos(\varpi_{m,l})=\frac{z_{l}-z_{m}}{D_{m,l}}$. The incidence angle at the mirror is obtained by projecting the unit vector pointing from the mirror back toward the AP onto the mirror normal
$\mathbf{n}_{m}$, which is given by
\vspace{-2mm}
{\small
\begin{equation}
\begin{aligned}
\cos(\beta_{m,l})={}&
\left(\frac{x_{l}-x_{m}}{D_{m,l}}\right)\sin(\varphi_{m})\cos(\vartheta_{m})
+\left(\frac{y_{l}-y_{m}}{D_{m,l}}\right)\\
&\cos(\varphi_{m})\cos(\vartheta_{m}) +\left(\frac{z_{l}-z_{m}}{D_{m,l}}\right)\sin(\vartheta_{m}).
\end{aligned}
\label{eq:inc_ap_mirror}
\end{equation}}

A departure angle $\varpi_{k,m}$ is obtained by projecting the unit vector pointing from the mirror toward the user onto the mirror normal, and is given by
\vspace{-1mm}
{\small
\begin{equation}
    \begin{aligned}
    \cos(\varpi_{k,m})={}&
\left(\frac{x_{k}-x_{m}}{D_{k,m}}\right)\sin(\varphi_{m})\cos(\vartheta_{m})
+\left(\frac{y_{k}-y_{m}}{D_{k,m}}\right)\\
&\cos(\varphi_{m})\cos(\vartheta_{m})
+\left(\frac{z_{k}-z_{m}}{D_{k,m}}\right)\sin(\vartheta_{m}).
\label{eq:irradiance_mirror}
    \end{aligned}
\end{equation}}

Since the reflected signal arrives at the user's photodetector, its incidence angle $\beta_{k,m}$ depends on the orientation of the user's device in the same way as the direct LoS link. It is measured at the receiver and therefore uses the unit vector pointing from the user toward the mirror, which is given by 

\vspace{-4mm}
{\small
\begin{equation}
\begin{aligned}
\cos(\beta_{k,m})={}&
\left(\frac{x_{m}-x_{k}}{D_{k,m}}\right)\cos(\varsigma)\sin(\Theta)
+\left(\frac{y_{m}-y_{k}}{D_{k,m}}\right)\\
&\sin(\varsigma)\sin(\Theta) +\left(\frac{z_{m}-z_{k}}{D_{k,m}}\right)\cos(\Theta).
\label{eq:inc_user_mirror}
\end{aligned}
\end{equation}}

Indoor optical links are frequently obstructed by people moving in the room. Each blocker is modeled as a vertical cylinder of radius $r_{b}$ and height $h_{b}$ moving independently over the floor under the same RWP model~\cite{chen2019omnidirectional}. A link is obstructed when a blocker intersects the segment joining its two endpoints. The same criterion is applied to each hop of the reflected path, so that a mirror contributes to a user's channel only when both the AP-mirror and mirror-user segments are clear. The blockage severity is quantified by the average fraction $\Upsilon$ of obstructed direct links, controlled through the number of blockers $N_{b}$, with $\Upsilon=0$ denoting the blockage-free case. Since the mirrors view each user from a range of directions, a blocker obstructing the direct link typically leaves part of the reflected paths clear, which the IRS can then steer toward the affected user.

\vspace*{-6mm}
\subsection{NOMA}
The transmitted optical signal at AP $l$ is constructed by superimposing the modulated symbols of all $K$ users in the power domain. The transmitted signal is expressed as \cite{maraqa2023optical}
\vspace{-1mm}
{\small  \begin{equation}
    x(t) = \sum_{j=1}^{K}\sqrt{P_j}s_j(t)+I_{\mathrm{DC}},
    \label{eq:tx}
\end{equation}}

\noindent where $P_{j}$ is the electrical power allocated to user $j$, $s_{j}(t)$ is the symbol information intended for user $j$, and $I_{\mathrm{DC}}$ is the DC bias current that ensures non-negative signals, as required in intensity-modulation direct-detection (IM/DD) VLC systems. The total electrical power $P_j$ available for allocation is bounded by $P_{\max}=(P_{t}/q)^{2}$, where $P_{t}$ is the optical transmit power and $q$ is the electrical-to-optical conversion coefficient. At the receiver side, the photodetector of user $k$ converts the received optical intensity into an electrical signal. After removing the DC component, the received signal of user $k$ is given by
\vspace{-1mm}
{\small
\begin{equation}
y_{k}=H_{k}\Big(\sum_{j=1}^{K}\sqrt{P_{j}}\,s_{j}\Big)+n_{k},
\label{eq:rx}
\end{equation}}

\noindent where $H_{k}=\sum_{l\in\mathcal{L}}H^{\mathrm{LoS}}_{k,l} +\sum_{m\in\mathcal{M}}\sum_{l\in\mathcal{L}}H^{\mathrm{IRS}}_{k,m,l}$ is the aggregate channel gain composed of the LoS and IRS-reflected signals of user $k$, and $n_{k}\sim\mathcal{N}(0,\sigma^{2})$ is the additive white Gaussian noise (AWGN), including both thermal and shot noise components, with variance $\sigma^{2}=BN_{0}$, where $B$ is the modulation bandwidth and $N_{0}$ is the noise power spectral density. In conventional NOMA, the SIC decoding order is typically predetermined based on the users' channel gains and assumed to be known at the receiver \cite{chen2017performance}. In static configurations, this ordering may be established using average channel gains and remain unchanged during operation. Consequently, it cannot adapt to time-varying channel conditions. In practical scenarios, however, user mobility and changes in receiver orientation lead to fluctuations in effective channel gains, potentially making a fixed decoding order inefficient and suboptimal for satisfying heterogeneous QoS requirements.

To overcome this limitation, we consider an adaptive decoding-order mechanism. By dynamically adjusting the orientation angles of the IRS mirrors, the effective channel gain of a target user can be selectively enhanced, thereby modifying its aggregate channel gain $H_k$ and potentially changing its relative position in the decoding order. Consequently, IRS reconfiguration enables the decoding order to adapt to the resulting effective channel conditions rather than remaining fixed according to the initial channel configuration. This capability is particularly beneficial in scenarios involving user mobility and variations in receiver orientation. Furthermore, IRS-assisted reflections can extend coverage to users located outside the direct illumination area of the AP. Accordingly, users are ordered in ascending order of their effective channel gains, such that the user with the weakest effective channel gain occupies the first decoding position, while the user with the strongest gain occupies the last. The power allocation is then optimized to satisfy the individual QoS requirements while ensuring successful SIC. Accordingly, the allocated powers must satisfy the following decoding-order constraint:
\vspace{-1mm}
{\small
\begin{equation}\label{eq:power_order}
P_{\Pi(i)} \geq P_{\Pi(i+1)}, \quad \forall i \in \{1, 2, \ldots, K-1\}~,
\end{equation}}

\noindent where $\Pi(i)$ denotes the index of the user occupying the $i$-th decoding position. Without loss of generality, the users are subsequently re-indexed according to their decoding order, such that user $k$ occupies the
$k$-th decoding position, i.e., $H_1 \leq H_2 \leq \cdots \leq H_K$. 
After performing SIC to cancel the signals of all previously decoded users, the received signal $y_k$ can be expressed as
\vspace{-2mm}
{\small
\begin{equation}\label{eq:sic_signal}
  y_k = H_k\sqrt{P_k}s_k +H_k\sum_{j=k+1}^{K}\sqrt{P_j}s_j +n_k.
\end{equation}}

\vspace*{-7mm}
\subsection{User Data Rate}
Following the decoding order established in the previous
subsection, the signal-to-interference-plus-noise ratio
(SINR) of user $k$ for decoding its own signal can be
expressed as
\vspace{-1mm}
{\small \begin{equation}
\Gamma_{k}=
\begin{cases}
\dfrac{(R_{p}H_{k})^{2}P_{k}}
{(R_{p}H_{k})^{2}\displaystyle\sum_{j=k+1}^{K}P_{j}
+BN_{0}}, & 1\leq k<K,\\[3mm]
\dfrac{(R_{p}H_{k})^{2}P_{k}}{BN_{0}}, & k=K,
\end{cases}
\label{eq:sinr}
\end{equation}}

\noindent where $R_p$ denotes the photodetector responsivity. Under the adopted decoding order, users are arranged in ascending order of their effective channel gains. Accordingly, user $K$, which has the strongest aggregate channel gain, can decode and cancel the signals of all preceding users before detecting its own signal, thereby eliminating
intra-cell interference under ideal SIC. Similarly, each user $k<K$ cancels the signals intended for users $1,\ldots,k-1$ and treats the signals of users $k+1,\ldots,K$ as interference. The achievable data rate of user $k$ is then given by
\cite{wang2013tight}
\vspace{-1mm}
{\small
\begin{equation}
R_{k}=B\log_{2}\left(1+
\frac{\exp(1)}{2\pi}\Gamma_{k}\right),
\label{eq:rate}
\end{equation}}

\noindent where $\exp(1)$ denotes Euler's number. This expression provides a lower bound on the channel capacity of IM/DD VLC systems. Finally, the aggregate system sum rate for the $K$ users is expressed as
\vspace{-2mm}
{\small
\begin{equation}
R_{T}=\sum_{k=1}^{K}R_{k}.
\label{eq:sumrate}
\end{equation}}

\vspace{-0.25in}
\section{Problem Formulation} \label{6.3}
An optimization problem is formulated to maximize the weighted system sum rate while satisfying the heterogeneous QoS requirements of all users. This is achieved by jointly optimizing the power allocation vector $\mathbf{P}$ and the IRS mirror orientation vectors $\boldsymbol{\varphi}$ and $\boldsymbol{\vartheta}$. By dynamically adjusting the IRS mirror orientations, the reflected signals can be steered to enhance the effective channel conditions of users according to their QoS requirements. Furthermore, the IRS-induced changes in the effective channel gains determine the decoding order. Accordingly, the optimization problem is formulated as follows:
\vspace{-1mm}
{\small
\begin{subequations}
\label{eq:P1_constraints_ieee}
\begin{align}
\textbf{P1:} \quad 
\max_{\mathbf{P}, \boldsymbol{\varphi}, \boldsymbol{\vartheta}} \quad 
& \sum_{k=1}^{K} w_k R_k \label{eq:P1_obj} \\
\text{s.t.} \quad 
& R_k \geq R_k^{\min}, \quad \forall k \in \mathcal{K}, \label{eq:P1_rate} \\
& P_{\Pi(i)} \geq P_{\Pi(i+1)}, \quad i = 1, \ldots, K-1, \label{eq:P1_sic_order} \\
& \sum_{k=1}^{K} P_k \leq P_{\max}, \label{eq:P1_power} \\
& P_k \geq 0, \quad \forall k \in \mathcal{K}, \label{eq:P1_nonneg} \\
& -\frac{\pi}{2} \leq \varphi_m \leq \frac{\pi}{2}, \quad \forall m \in \mathcal{M}, \label{eq:P1_phi} \\
& -\frac{\pi}{2} \leq \vartheta_m \leq \frac{\pi}{2}, \quad \forall m \in \mathcal{M}, \label{eq:P1_theta}
\end{align}
\end{subequations}}

\noindent where $w_k$  denotes the weight assigned to user $k$ to reflect its priority in the weighted sum-rate objective with $\sum_{k=1}^{K} w_k = 1$, $\mathbf{P} = [P_1, \ldots, P_K]^T$ is the power allocation vector, and ${\Pi(i)}$ represents the index of the user occupying the $i$-th decoding position. The vectors $\boldsymbol{\varphi} = [\varphi_1, \ldots, \varphi_M]^T$ and $\boldsymbol{\vartheta} = [\vartheta_1, \ldots, \vartheta_M]^T$ are the IRS roll and yaw angle of the $M$ mirror elements, respectively. Each user achievable rate $R_k$ depends on $\mathbf{P}$, $\boldsymbol{\varphi}$, and $\boldsymbol{\vartheta}$ through the resulting effective channel gains. Since the IRS mirror orientations directly reshape the aggregate channel gains $H_k$, the decoding order is induced by sorting the resulting effective channel gains in ascending order as
\vspace{-1mm}
{\small
\begin{equation}
\Pi=\operatorname{argsort}\big(\{H_{k}^{2}\}_{k=1}^{K}\big),\qquad
H_{\Pi(1)}\leq H_{\Pi(2)}\leq\cdots\leq H_{\Pi(K)}.
\label{eq:sort}
\end{equation}}

Accordingly, constraint~\eqref{eq:P1_rate} ensures the minimum QoS requirement of each user, while constraint~\eqref{eq:P1_sic_order} enforces the power allocation follows the decoding order, so that users with the weaker effective channel gain are allocated higher transmission powers. Constraint~\eqref{eq:P1_power} ensures that the total allocated power does not exceed the maximum budget $P_{\max}$, while constraint~\eqref{eq:P1_nonneg} enforces a non-negative power allocation. Constraints~\eqref{eq:P1_phi} and~\eqref{eq:P1_theta} define the feasible orientation ranges of IRS mirrors, respectively. Since the channel gains that induce the decoding-order permutation $\Pi$ depend on the optimization variables, the feasible set associated with ~\eqref{eq:P1_sic_order} is variable-dependent, which is the combinatorial structure that makes \textbf{P1} non-convex coupling between the variables. The problem is therefore NP-hard, with a search space growing exponentially with the number of users $K$ and IRS mirror elements $2M$. Consequently, solving P1 using conventional optimization techniques can be computationally demanding, particularly in dynamic environments where user positions, receiver orientations, and traffic demands vary over time.
To overcome these challenges, we propose an intelligent DNN-DRL framework for real-time resource allocation. First, a DNN estimates the users' positions and orientations from the available received-signal measurements. Based on these estimates and the resulting system state, the DRL agent jointly optimizes the user power allocation and IRS mirror orientations, from which the decoding order follows, allowing it to adapt dynamically to changes in the users' channel conditions. The proposed framework is detailed in the next section.

\vspace{-3mm}
\section{Adaptive Framework for Dynamic Scenarios} \label{6.4}
Applying conventional DRL directly to the high-dimensional action space of~\eqref{eq:P1_constraints_ieee} leads to slow convergence and reduced solution accuracy, as the agent must explore an exponentially expanding set of joint configurations to identify high-reward actions. This challenge motivates the adoption of a more structured learning framework. Accordingly, the proposed two-stage architecture uses DNN-based position and orientation estimation to provide the agent with accurate user-state information so that the agent does not need to infer the user positions from raw observations and can focus on the resource allocation task. This reduces the learning burden on the DRL agent and improves the overall learning efficiency. The two-stage framework is detailed in the following subsections.
 \vspace{-5mm}\subsection{DNN-Based User Position and Orientation Estimation}
The performance of the proposed IRS-assisted NOMA-OWC network depends critically on accurate knowledge of each user's position and orientation, which, together with the IRS configuration, determine the aggregate channel gains $H_k$ and consequently influence the decoding order and achievable data rates. In dynamic indoor environments, user mobility and changes in receiver orientation cause the channel conditions to vary over time. Consequently, frequent explicit channel estimation may introduce substantial pilot and signalling overhead. To address this challenge, a DNN is employed to estimate the three-dimensional positions and orientations of each user using received signal strength (RSS) measurements of IR uplink signals collected at the APs~\cite{arfaoui2021invoking}. The estimated user states are subsequently used to determine the corresponding downlink channel gains, thereby supporting the DRL-based resource allocation process.

\vspace{-1mm}
\subsubsection{Data Generation}
During the offline training phase, each user transmits an upload IR reference signal with optical transmit power $P_{\mathrm{IR}}$. The received optical power $P_{k,l}^{\mathrm{rx}}$ at the $l$-th AP from user $k$ is expressed as
\vspace{-1mm}
{\small
\begin{equation}
P_{k,l}^{\mathrm{rx}}
=P_{\mathrm{IR}} H_{k,l}^{\mathrm{UL}},
\label{eq:received_IR_power}
\end{equation}}

\noindent where $H_{k,l}^{\mathrm{UL}}$ denotes the corresponding uplink IR channel gain. To investigate the contribution of IRS-reflected paths to user-state estimation, two propagation configurations are considered.
In the first configuration, only the direct uplink LoS channel gain component $H_{k,l}^{\mathrm{LoS,IR}}$ is considered. In the second configuration, both uplink direct LoS  $H_{k,l}^{\mathrm{IRS,IR}}$and IRS-reflected IR channel gain $H_{k,m,l}^{\mathrm{IRS,IR}}$ are considered. 
Note that, the IRS orientation is kept fixed during DNN training. This ensures that the RSS fingerprints are generated under a consistent IRS configuration.

\vspace{-1mm}
\subsubsection{DNN Architecture}
For each user, the RSS measurements collected at the $L$ APs form the input fingerprint to the DNN. The network learns the nonlinear mapping between the RSS fingerprint and the user's position and orientation, i.e.,  $\mathbf{e}_{k} = [x_k,y_k,z_k,\varsigma_k,\Theta_k]^{T},$
where $(x_k,y_k,z_k)$ denotes the three-dimensional user position, while $\varsigma_k$ and $\Theta_k$ represent the user's orientation angles. The DNN consists of an input layer, $F_h$ fully connected hidden layers, and an output layer. The input layer $\mathbf{u}_{0}=\mathbf{r}_{k}^{c},$ where $c\in\{LoS,{LoS-IRS}\}$ identifies the propagation configuration. The output of each hidden layer is calculated as
{\small
\begin{equation}
\mathbf{u}_{j}
= \operatorname{ReLU} \left(
\mathbf{W}_{j}\mathbf{u}_{j-1} + \mathbf{b}_{j}
\right),
\quad j=1,\ldots,F_h,
\label{eq:DNN_hidden}
\end{equation}}

\noindent where $\mathbf{W}_{j}$ and $\mathbf{b}_{j}$ denote the weight matrix and bias vector of the $j$-th hidden layer, respectively. The output layer estimates the user's position and orientation according to
{\small
\begin{equation}
\hat{\mathbf{e}}_{k} = \mathbf{W}_{\mathrm{out}}\mathbf{u}_{F_h} + \mathbf{b}_{\mathrm{out}},
\label{eq:DNN_output}
\end{equation}}

\noindent where $\hat{\mathbf{e}}_{k}
[\hat{x}_k,\hat{y}_k,\hat{z}_k,\hat{\varsigma}_k,\hat{\Theta}_k]^{T}$ is the estimated user-state.

\vspace{-2mm}
\subsubsection{DNN Training and Estimation}
The DNN is trained using a dataset containing user-state samples and their corresponding RSS fingerprints. Each sample consists of a known user position and orientation, together with the RSS measurements collected at the APs under the selected propagation configuration. The dataset is
divided into training, validation, and testing subsets. Before training, the RSS input features and output labels are normalized to improve numerical stability and facilitate convergence. The network parameters are optimized by minimizing the mean squared error (MSE)
between the normalized ground-truth user states and the  corresponding DNN estimates given by
{\small
\begin{equation}
\boldsymbol{\theta}^{*} = \arg\min_{\boldsymbol{\theta}}
\frac{1}{N_{\mathrm{train}}} \sum_{n=1}^{N_{\mathrm{train}}}
\left\| \tilde{\mathbf{e}}^{(n)} - \hat{\tilde{\mathbf{e}}}^{(n)}
\left( \mathbf{r}^{c,(n)};\boldsymbol{\theta} \right) \right\|_{2}^{2},
\label{eq:DNN_loss}
\end{equation}}

\noindent where $N_{\mathrm{train}}$ is the number of training samples, $\boldsymbol{\theta}$ represents the trainable network parameters, and the tilde denotes normalized quantities. The Adam optimizer is employed
to update the network parameters, while the validation subset is used to monitor training performance and select the final model. During online operation, the controller monitors the RSS measurements of the received IR uplink signals at the APs. Whenever significant changes are detected, the updated RSS fingerprint $\mathbf{r}_{k}^{c}$ is fed into the trained DNN to obtain an updated estimate of the user's position and orientation, $\hat{\mathbf{e}}_{k} = f_{\mathrm{DNN}} \left(\mathbf{r}_{k}^{c}\right),$ where $f_{\mathrm{DNN}}(\cdot)$ denotes the trained DNN mapping.  
The estimated user states are subsequently forwarded to the DRL agent. In particular, the estimated positions and orientations are used to calculate the VLC downlink channel gains. Based on these channel estimates, the DRL agents optimize the transmit power allocation and IRS orientation angles to improve system performance while satisfying the considered QoS and power constraints. The DNN is evaluated using an unseen test dataset, 
containing $N_{\mathrm{test}}$ samples. The position estimation accuracy, $A_{\chi}$, is defined as the fraction of test samples for which the Euclidean position estimation error does not exceed a predefined threshold $\chi$, given by
\vspace{-1mm}
{\small
\begin{equation}
A_{\chi} = \frac{1}{N_{\mathrm{test}}} \sum_{n=1}^{N_{\mathrm{test}}} \mathbbm{1} \left( \left\|
\hat{\mathbf{L}}_{n} - \mathbf{L}_{n} \right\|_{2} \leq \chi
\right),\label{eq:accuracy}
\end{equation}}

\noindent where $\mathbbm{1}(\cdot)$ denotes the indicator function, $\hat{\mathbf{L}}_{n}$ and $\mathbf{L}_{n}$ represent the estimated and ground-truth three-dimensional positions of the user in the $n$-th test sample, respectively, and $\chi$ is the predefined position error threshold.
\begin{figure*}[t]
\centering
\includegraphics[width=0.7\textwidth]{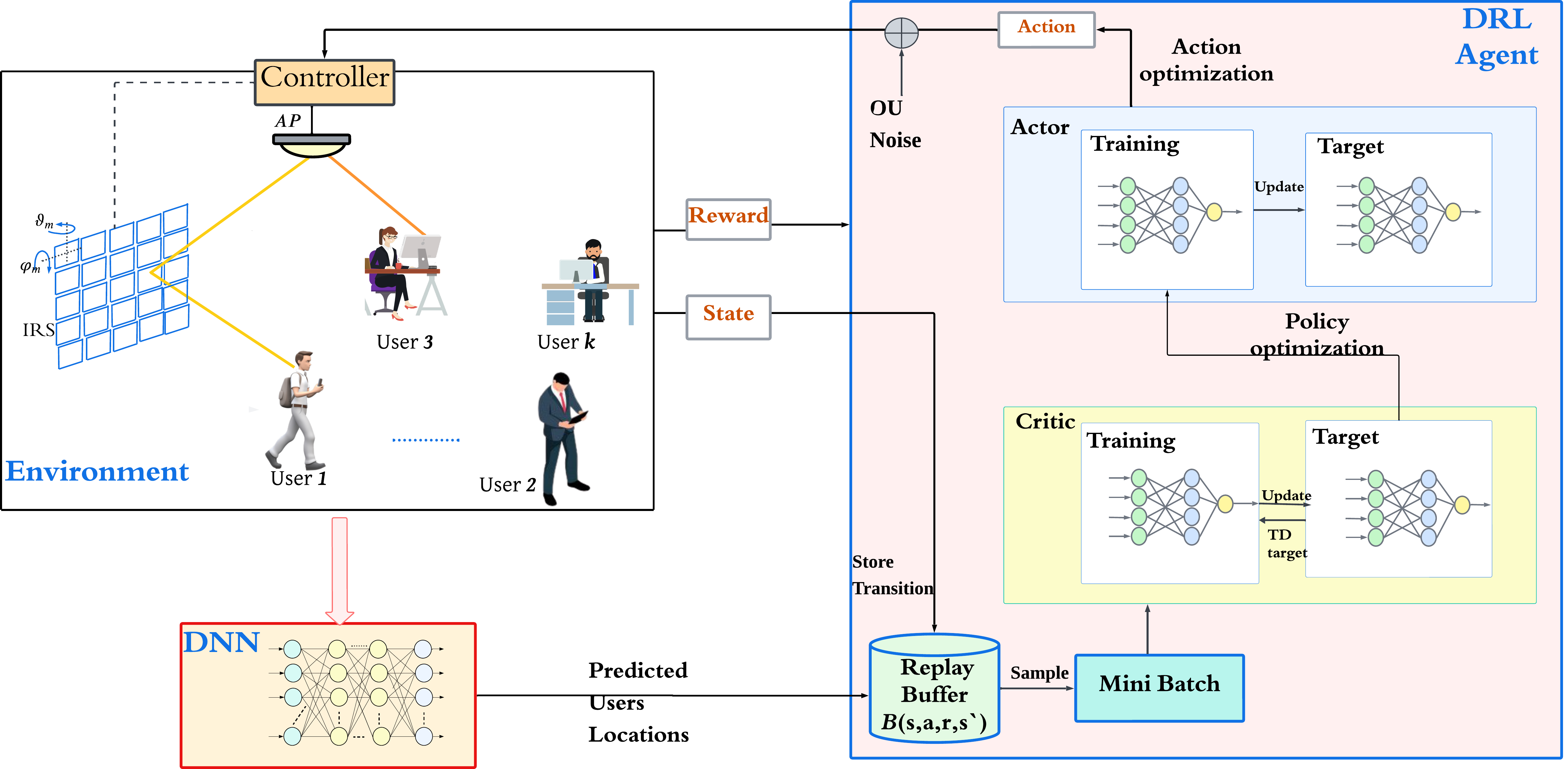}
\caption{The proposed DNN-DRL architecture. }
\label{fig:DDPG_arch}
\vspace{-5mm}
\end{figure*}
\vspace*{-4mm}
\subsection{Intelligent DRL Algorithm for Solving Optimization} 

The second stage adopts deterministic policy-based DRL algorithm to jointly optimize the continuous power allocation and IRS orientation variables.  To this end, the optimization problem \textbf{P1} is reformulated as a MDP, allowing the agent to learn an effective control policy through continuous interaction with the environment. The MDP formulation and the implementation of the proposed DRL algorithm are detailed in the following.

\vspace{-2mm}\subsubsection{MDP Formulation}

The optimization problem \textbf{P1} is reformulated as an MDP defined by the tuple $\langle \mathcal{S}, \mathcal{A}, \mathcal{P}, \mathcal{R}, \gamma \rangle$, where $\mathcal{S}$ is the state space, $\mathcal{A}$ is the action space, $\mathcal{P}$ is the state transition probability, $\mathcal{R}$ is the reward function designed to reflect the objective of \textbf{P1}, and $\gamma \in (0,1)$ is the discount factor. The environment represents the indoor IRS-assisted NOMA-VLC system, comprising the AP transmitters, IRS mirrors, VLC channel, and mobile users. The DRL agent is located at the central controller and interacts with the environment to learn a policy $\pi$ that maximizes the cumulative expected reward given by
\vspace{-2mm}
{\small
\begin{equation}
\max_{\pi}\ \mathbb{E}_{\pi}\!\left[\sum_{t=0}^{T}\gamma^{t} r_{t}\right],
\label{eq:cumreward}
\end{equation}}

\noindent where $r_{t}$ is the reward at time step $t$, and $T$ is the total number of
time steps per episode.
 
\textit{State space:} The state space $\mathcal{S}$ consists of all possible system states $s_{t} \in \mathcal{S}$, where $s_{t}$ is the state at time step $t$. The state includes the user positions $\hat{\mathbf{L}}_t$ and orientations $\hat{\mathbf{O}}_t$ estimated by the first-stage DNN, the users' minimum-rate requirement vector $\mathbf{R}^{\min}=[R^{\min}_1,\dots, R^{\min}_K]^{T}$, the power allocation $\mathbf{P}_{t-1}$ deployed at the previous time step, and the previous IRS roll $\bm{\varphi}_{t-1}$ and yaw $\bm{\vartheta}_{t-1}$ angles. 
The decoding order is derived from the effective channel gains of the users. Specifically, after selecting an action and determining the corresponding IRS configuration, the agent evaluates the resulting aggregate channel gains and sorts them in ascending order to obtain $\Pi_{t}$. 
 
\textit{Action space:} The action space $\mathcal{A}$ comprises the continuous control variables selected by the agent at each time step $t$. Given the observed state $s_t$, the agent selects an action $a_t \in \mathcal{A}$ consisting of the power allocation for the $K$ users and the roll $\bm{\varphi}_{t}$ and yaw $\bm{\vartheta}_{t}$ of the $M$ IRS mirrors, expressed as
\vspace{-2mm}
{\small
\begin{equation}
a_t = \left[ \tilde{\mathbf{P}}_t^T,
\bm{\varphi}_t^T, \bm{\vartheta}_t^T
\right]^T \in \mathbb{R}^{K+2M}, \label{eq:action}
\end{equation}}

\noindent where ${\tilde{\mathbf{P}}_t}$ denotes the power-allocation vector generated by the actor network. The resulting power allocation is mapped to a feasible non-negative vector satisfying the total power constraint in~\eqref{eq:P1_power}. Meanwhile, the IRS angles are constrained to the interval $[-\pi/2,\pi/2]$, in accordance with \eqref{eq:P1_phi} and \eqref{eq:P1_theta}.

\textit{Reward function:} The reward function $\mathcal{R}$ is designed to align with the objective of \textbf{P1} while guiding the agent to satisfy the minimum-rate requirements of all users. At each step, the agent receives an immediate reward consisting of the weighted sum rate and a penalty for QoS violations, expressed as
\vspace{-2mm}
{\small
\begin{equation}
r_{t}=\sum_{k=1}^{K} w_{k} R_{k,t}
-\lambda \sum_{k=1}^{K}\big[\max\!\left(0,\ R^{\min}_k-R_{k,t}\right)\big]^{2},
\label{eq:reward}
\end{equation}}
\vspace{-1mm}

\noindent where $R_{k,t}$ is the achievable rate of user $k$ at step $t$, $w_k$ is its corresponding weighting coefficient, $\lambda>0$ is a penalty coefficient that controls the QoS satisfaction. 

\vspace{-4mm}\subsubsection{DPG-based DRL Algorithm}
While DRL is a good solution for sequential decision-making in a dynamic environment, standard DRL algorithms cannot directly and efficiently solve the problem in \eqref{eq:P1_constraints_ieee}. A standard DRL algorithm learns a policy by mapping observed states to actions through repeated interaction with the environment, and the difficulty of learning this mapping grows sharply with the size of the action space. The continuous and high-dimensional action space of \textbf{P1}  makes value-based methods such as DQN unsuitable, since discretizing the action set introduces approximation errors and yields a number of actions that grows exponentially with $K$ and $M$.
For this reason, a DPG-based DRL algorithm is designed for the continuous control of the IRS-assisted NOMA-OWC system. The agent jointly optimizes the power allocation $\mathbf{P}_t$ and the IRS roll and yaw angles $(\boldsymbol{\varphi}_t, \boldsymbol{\vartheta}_t)$, from which the SIC decoding order $\boldsymbol{\Pi_t}$ emerges through the IRS-configured channel gains. By inducing the decoding order rather than treating it as an explicit discrete action, the algorithm keeps a purely continuous action space and avoids the hybrid-action formulation that limits conventional DRL methods. The proposed algorithm follows an actor-critic architecture comprising four neural networks: actor-network $\mu$ and critic-network $Q^{\mu}$, which are called the evaluation networks, together with their corresponding target networks $\mu'$ and $Q^{\mu'}$, as shown in Fig. \ref{fig:DDPG_arch}. These networks work together to learn an effective QoS-aware resource allocation policy. It is worth mentioning that the target networks have the same architecture as the evaluation network and are used to help stabilize training and to prevent rapid changes in value estimation, which can cause instability.  The DPG framework learns a deterministic policy that maps each state directly to a continuous action, maximizing the expected return without action discretization \cite{silver2014deterministic}. In the following, the network architectures and their roles are described in detail. 

\begin{itemize}[leftmargin=*,nosep]
\item \textbf{Actor Network} $\mu(s|\boldsymbol{\theta}^\mu): \mathcal{S} \to \mathcal{A}$: learns the deterministic policy that maps states to continuous actions to maximize the expected reward. It is parameterized by a deep neural network with weights $\boldsymbol{\theta}^\mu$, and the selected action given by 
\vspace{-2mm}
\begin{equation}
a_t = \mu(s_t|\theta^\mu).
\label{eq6:actor}
\end{equation}
The objective of the training is to learn a policy that maximizes the expected cumulative return given by
\vspace{-1mm}
{\small
\begin{equation}
\mathcal{J}(\theta^\mu) = \mathbb{E}_{s_t \sim \mathcal{D}}\!\left[ Q(s_t,\mu(s_t|\theta^\mu)|\theta^Q) \right].
\label{eq6:policyGradient}
\end{equation}}

By using the DPG, which encourages actions to maximize $Q(s_t,a_t)$, the actor network parameters $\theta^{\mu}$ are trained and updated using policy gradient $\nabla_{\theta^\mu} \mathcal{J} $, given by 
\vspace{-1mm}
{\small
\begin{equation}
\label{eq:policyGradient}
\nabla_{\theta^\mu} \mathcal{J} = \frac{1}{\mathcal{N}_b} \sum_{i=1}^{\mathcal{N}_b}
\nabla_a Q(s_i, a | \theta^{Q}) \big|_{a = \mu(s_i | \theta^\mu)} 
\nabla_{\theta^\mu} \mu(s_i | \theta^\mu),
\end{equation}}

\noindent where the first term $\nabla_a Q\left(s, a\ \middle|\theta^{Q}\right)$ is the gradient of the action-value function $Q(s, a)$  with respect to the action $a$ evaluated at the actor's output $a = \mu(s|\theta^\mu)$. This gradient indicates how a change in the action would affect the expected future return. The second term $\nabla_{\theta_\mu} \mu\left(s\middle|\theta^\mu\right)$ is the gradient of the deterministic policy with respect to its parameters $\theta^\mu$. This represents the direction in which the policy is adjusted for improvement. The gradient is backpropagated through the critic using the actions produced by the actor.

\item \textbf{Critic Network} 
$Q(s,a|\boldsymbol{\theta}^Q): \mathcal{S} \times \mathcal{A} \to \mathbb{R}$ estimates the action-value function (Q-function), parameterized by weights $\boldsymbol{\theta}^Q$. It evaluates the expected cumulative reward obtained by taking action $a$ in state $s$ and following the current policy represented as
 \vspace{-4mm}

{\small
\begin{align}
Q(s_t,a_t|\boldsymbol{\theta}^Q) &= \mathbb{E}\!\left[\sum_{\tau=t}^{\infty} \gamma^{\tau-t} r_\tau \,\Big|\, s_t, a_t\right], \label{eq:q_function}
\end{align}}

\noindent where $\gamma \in (0,1)$ is the discount factor balancing immediate and future rewards. Following the Bellman equation, the target value $y_i$ is computed using the target networks as
\vspace{-3mm}
{\small
\begin{equation}
y_i = r_i + \gamma\, Q'\!\big(s_{i+1}, \mu'(s_{i+1}|\theta^{\mu'}) \,\big|\, \theta^{Q'}\big),
\label{eq6:y_i}
\end{equation}}

\noindent where $Q'(\cdot|\theta^{Q'})$ is the target critic network that estimates the value of the next state-action pair, $\mu'(\cdot|\theta^{\mu'})$ is the target actor network that produces the next action $\mu'(s_{i+1}|\theta^{\mu'})$, and $\theta^{\mu'}$ and $\theta^{Q'}$ are the target actor and target critic parameters, respectively. The critic parameter $\theta^{Q}$ is updated to improve its accuracy in predicting the expected future reward for a given state-action pair by minimizing the mean-squared-error (MSE) loss $L(\theta^Q)$ between the predicted and target Q-values, expressed as
\vspace{-2mm}

{\small
\begin{equation} 
L(\theta^Q) = \frac{1}{\mathcal{N}_b} \sum_{i=1}^{\mathcal{N}_b} \left( y_i - Q(s_i, a_i| \theta^Q) \right)^2,
\label{eq:Loss}
\end{equation}}

\noindent where $Q(s_i, a_i | \theta^Q)$ is the predicted Q-value. Minimizing this loss aligns the critic's predictions with the target values, improving the accuracy of the value estimation, which is in turn used to guide the actor update.

\item \textbf{Target Networks} $\mu'(s|\boldsymbol{\theta}^{\mu'})$ and 
$Q'(s,a|\boldsymbol{\theta}^{Q'})$: stabilize training by preventing rapid changes in the value estimates that would otherwise cause instability. Both networks share the same architecture as the evaluation networks and are used to compute the learning targets. The target network parameters are adjusted gradually via soft updates as follows
\vspace{-2mm}
{\small
\begin{equation}
\boldsymbol{\theta}^{\mu'}
\leftarrow \tau \boldsymbol{\theta}^{\mu}
+ (1-\tau)\boldsymbol{\theta}^{\mu'},
\quad
\boldsymbol{\theta}^{Q'}
\leftarrow \tau \boldsymbol{\theta}^{Q}
+ (1-\tau)\boldsymbol{\theta}^{Q'},
\label{eq:soft_update}
\end{equation}
}

\noindent where $\tau \ll 1$ is the soft-update coefficient that ensures slow target updates for stability. An experience replay memory $\mathcal{D}$ is used to store all experience transitions $\{s_{t}, a_{t}, r_{t}, s_{t+1}\}$ encountered during training. During each update, a random mini-batch of size $\mathcal{N}_b$ is sampled from $\mathcal{D}$, which reduces the correlation among the samples and improves learning stability~\cite{ahrarICC}. 

\item \textbf{Exploration}: 
Since the deterministic policy provides limited exploration in continuous action spaces, an Ornstein-Uhlenbeck (OU) process is employed to generate temporally correlated exploration noise. This correlated noise enables smoother exploration and helps the policy avoid local optima. The exploratory action is obtained by adding the noise to the deterministic action, which can be written as~\cite{wang2023ddpg}
\vspace{-3mm}
{\small
\begin{equation}
a_t = \mu(s_t|\theta^\mu) + \mathcal{X}_t,
\label{eq:Actornoise}
\end{equation}}

\noindent where $\mathcal{X}_t$ denotes the OU exploration noise. Each executed action yields a reward $r_t$ and a next state $s_{t+1}$, which together with $s_t$ and $a_t$ are stored as a transition in the replay buffer $\mathcal{D}$.
\end{itemize}

\vspace*{-2mm}
\subsubsection{Algorithm description}
The proposed DPG-based DRL algorithm is summarized in Algorithm~\ref{alg:proposed}. The actor and critic networks, their corresponding target networks, and the replay buffer $\mathcal{D}$ are initialized in steps~1-2. The training process consists of $N_{\mathrm{ep}}$ episodes, each containing $T$ time steps. At the beginning of each episode, the environment and the OU exploration noise process are initialized. The agent receives the DNN-estimated user positions and orientations, the minimum-rate requirements, and the previously deployed power allocation and IRS orientations, which are used to construct the initial state $s_0$. At each time step, the actor generates a continuous action. The action variables are subsequently mapped to their feasible physical ranges, with the power allocation normalized to satisfy the total power constraint in~\eqref{eq:P1_power}. The agent then applies the selected IRS orientations and evaluates the resulting effective channel gains. Based on these gains, the decoding order $\Pi_t$ is determined by sorting the users in ascending order of their effective channel gains, as specified in~\eqref{eq:sort}. Thus, the decoding order is induced by the resulting channel conditions rather than being an independent decision variable of the actor.
Next, the candidate power values are assigned to the users according to $\Pi_t$ in descending order of power magnitude. Consequently, users with weaker effective channel gains receive higher power allocations, thereby satisfying the power-ordering constraint in~\eqref{eq:P1_sic_order}. The resulting power allocation and decoding order are then used to calculate the users' achievable rates and the immediate reward according to~\eqref{eq:reward}. The agent observes the next state $s_{t+1}$, and the resulting transition is stored in the replay buffer. In accordance with the implemented action mapping, the replay buffer stores the raw actor action before the environment-side power normalization and power-ordering projection. The critic therefore takes the raw action as its input and learns its corresponding value after the deterministic action transformations performed by the environment. Once the replay buffer contains at least $N_b$ transitions, a random mini-batch is sampled to update the critic and actor networks. The target networks are subsequently updated using the soft-update rule in~\eqref{eq:soft_update}. By dynamically adjusting the IRS orientations, the proposed framework reshapes the users' effective channel gains and consequently adapts the decoding order to instantaneous channel conditions. The associated power allocation is jointly optimized to maximize the weighted sum rate while encouraging the satisfaction of the users' minimum-rate requirements.

\makeatletter
{\setlength{\@fptop}{0.75in}
\begin{algorithm}[t]
\small
\caption{Proposed DRL Algorithm for Joint Power Allocation and IRS Orientation}
\label{alg:proposed}
\begin{algorithmic}[1]
\setlength{\baselineskip}{0.95\baselineskip}
\Require Number of episodes $N_{\mathrm{ep}}$, steps per episode $T$, actor and
critic learning rates $\mathcal{E}_{a}$, $\mathcal{E}_{c}$
\State \textbf{Objective:} Maximize the QoS-weighted sum rate of the network
\State \textbf{Initialize:} actor network $\mu(s|\theta^{\mu})$ and critic
network $Q(s,a|\theta^{Q})$ with random weights $\theta^{\mu}$, $\theta^{Q}$;
target networks $\mu'(s|\theta^{\mu'})$ and $Q'(s,a|\theta^{Q'})$ with
$\theta^{\mu'}\!\leftarrow\!\theta^{\mu}$,
$\theta^{Q'}\!\leftarrow\!\theta^{Q}$; replay buffer $\mathcal{D}$, mini-batch
size $N_{b}$, discount factor $\gamma$, soft-update rate $\tau$
\For{episode $=1$ \textbf{to} $N_{\mathrm{ep}}$}
    \State Initialize the environment and a random OU noise process $X_{t}$ 
    \State Receive the DNN-estimated user positions and orientations,
           $\mathbf{R}^{\min}$, and the power allocation and
           IRS angles; form the initial state $s_{0}$
        \For{$t=0$ \textbf{to} $T-1$}
        \State Select action $a_{t}=\mu(s_{t}|\theta^{\mu})+X_{t}$
        \State Extract the power vector $\tilde{\mathbf{P}}_{t}$ and the IRS
               angles $(\bm{\varphi}_{t},\bm{\vartheta}_{t})$ from $a_{t}$;
               rescale to their physical ranges and normalize
               $\tilde{\mathbf{P}}_{t}$ to satisfy \eqref{eq:P1_power}
        \State Apply the IRS orientations
               $(\bm{\varphi}_{t},\bm{\vartheta}_{t})$ and evaluate the
               aggregate channel gains
               $\{H_{k}(\bm{\varphi}_{t},\bm{\vartheta}_{t})\}$ from
               \eqref{eq:losch}--\eqref{eq:inc_user_mirror}
        \State Induce the decoding order $\Pi_{t}$ by sorting $\{H_{k}^{2}\}$ in
               ascending order using \eqref{eq:sort}
        \State Project $\tilde{\mathbf{P}}_{t}$ onto $\Pi_{t}$: assign its
               entries in descending magnitude to the users in the order
               $\Pi_{t}$ to obtain $\mathbf{P}_{t}$, enforcing
               \eqref{eq:P1_sic_order}
        \State Compute the user rates $\{R_{k,t}\}$ under $\Pi_{t}$
               and the reward $r_{t}$ from \eqref{eq:reward}; observe the next
               state $s_{t+1}$
        \State Store the transition $(s_{t},a_{t},r_{t},s_{t+1})$ in
               $\mathcal{D}$
        \If{$|\mathcal{D}| \geq N_{b}$}
            \State Sample a random mini-batch of $N_{b}$ transitions
                   $(s_{i},a_{i},r_{i},s_{i+1})$ from $\mathcal{D}$
            \State Compute the target $Q$-values $y_{i}$ using \eqref{eq6:y_i}
            \State Update the critic $\theta^{Q}$ by minimizing loss in \eqref{eq:Loss}
            \State Update the actor $\theta^{\mu}$ using policy gradient in \eqref{eq:policyGradient}
            \State Soft-update target networks $\theta^{\mu'}$,
                   $\theta^{Q'}$ using \eqref{eq:soft_update}
        \EndIf
        \State $s_{t} \leftarrow s_{t+1}$
    \EndFor
\EndFor
\vspace{-1mm}
\end{algorithmic}
\end{algorithm}}
\makeatother

\vspace{-2mm}
\subsection{Computational Complexity}
The computational complexity of the proposed framework comprises online inference
and offline training. In the first stage, the DNN estimates the position and
orientation of each user from its IR uplink SNR fingerprint, at a cost of
$\mathcal{O}\big(K\sum_i n_i^d n_{i+1}^d\big)$, where $n_i^d$ denotes the number
of neurons in layer $i$ of the DNN and the network is evaluated once per user. In
the second stage, the actor forward pass requires
$\mathcal{O}\big(\sum_i n_i^a n_{i+1}^a\big)$ operations, where $n_i^a$ denotes
the actor layer widths. Evaluating the aggregate channel gains for $K$ users and
$M$ IRS mirrors requires $\mathcal{O}(KM)$, determining the decoding order and
projecting the power allocation require $\mathcal{O}(K\log K)$, and the achievable
rates are obtained in $\mathcal{O}(K)$ using cumulative sums to evaluate the NOMA
interference terms. The overall online complexity is therefore
$\mathcal{O}\big(\sum_i n_i^a n_{i+1}^a + KM + K\log K\big)$, which is dominated
by the actor forward pass. The decoding order is thus induced through sorting,
avoiding an exhaustive search over the $K!$ possible user permutations. Training
is performed offline using experience replay and, assuming one update per
environment step, has a complexity of approximately
$\mathcal{O}\big(N_{\mathrm{ep}}TN_b(C_{\mu}+C_Q)\big)$, where $C_{\mu}$ and $C_Q$
denote the forward-pass complexities of the actor and critic networks and $N_b$ is
the mini-batch size. For comparison, DQN requires discretizing the continuous action space, which
yields $G_a^{K+2M}$ possible actions for $G_a$ levels per dimension, whereas the DPG-based algorithm generates continuous control actions directly. The SCA benchmark has an approximate complexity of
$\mathcal{O}\big(N_{\mathrm{it}}(K+2M)^3\big)$ under a generic dense solver. The exhaustive-search benchmark evaluates $G_P^K$ power configurations, and for each
of them the additive structure of~\eqref{IRSCH} allows every mirror orientation to be optimized separately over a $G\times G$ angular grid, giving
$\mathcal{O}\big(G_P^K M G^2\big)$ rather than $\mathcal{O}\big(G_P^K G^{2M}\big)$. It remains exponential in $K$ and is therefore used as an offline upper bound only.

\begin{table}[b]
\centering
\caption{System and Learning Parameters.}
\label{tab:learning_params}
\scriptsize
\setlength{\tabcolsep}{2pt}
\renewcommand{\arraystretch}{1}
\begin{tabular}{@{}lc|lc@{}}
\hline
\multicolumn{2}{c|}{\textbf{System Parameters}} &
\multicolumn{2}{c}{\textbf{Learning Parameters}} \\
\cline{1-2} \cline{3-4}
\textbf{Parameter} & \textbf{Value} &
\textbf{Parameter} & \textbf{Value} \\
\hline
LED half-power angle $\Phi_{1/2}$ & $60^\circ$ &
Actor learning rate $\mathcal{E}_a$ & $10^{-4}$ \\
PD area $A_r$ & $20~\mathrm{mm}^2$ &
Critic learning rate $\mathcal{E}_c$ & $10^{-3}$ \\
Concentrator gain $g(\psi_c)$ & 1.5 &
Discount factor $\gamma$ & 0.99 \\
Photodetector responsivity  $R_p$ & 0.5 A/W &
Replay buffer $\mathcal{D}$ & $10^6$ \\
Bandwidth $B$ & $20~\mathrm{MHz}$ &
Soft-update factor $\tau$ & $10^{-3}$ \\
Noise PSD $N_0$ & $10^{-21}~\mathrm{A}^2/\mathrm{Hz}$ &
Mini-batch size $\mathcal{N}_b$ & 64 \\
\hline
\end{tabular}
\vspace{-3mm}
\end{table}

\vspace*{-2mm}
\section{Simulation and Performance Evaluation} \label{sec:6.5}
In this section, the performance of the proposed framework is evaluated by solving the formulated optimization problem within an indoor IRS-assisted NOMA-OWC environment.
\vspace{-2mm}
\subsection{System Configuration}
We consider the indoor IRS-assisted NOMA-OWC network described in Section~\ref{sec:sm}. The room dimensions are $5\,\mathrm{m}\times5\,\mathrm{m}\times3\,\mathrm{m}$, with eight optical APs ($L=8$) mounted on the ceiling. Unless otherwise specified, each AP transmits an optical power of $P_t=2$~W. To maintain connectivity under user mobility and blockage, an IRS comprising a $7\times7$ array of reflective mirror elements ($M=49$) is mounted on one wall. Each mirror has a reflection coefficient of $0.95$ and dimensions of $25\,\mathrm{cm}\times10\,\mathrm{cm}$, with a $10$~cm space between adjacent elements. The receiving plane accommodates $K=5$ randomly distributed users, with minimum-rate requirements $\mathbf{R}^{\min}=[5,10,10,15,15]$~Mbps, representing different traffic requirements. The blockers have a radius of $r_b=0.15$~m and a height of $h_b=1.75$~m, while the user receiver plane is located at $h_u=0.85$~m. Users move at randomly selected speeds in the range $[0,2]$~m/s, with zero pause time. The remaining system parameters are listed in Table~\ref{tab:learning_params}.

\vspace{-1mm}
\subsection{Training Configuration}
The simulation environment is implemented using Python 3.12. The DNN takes an eight-dimensional IR uplink RSS fingerprint, corresponding to the received contributions at the $L=8$ APs, as its input. A single shared-weight DNN is deployed at the central controller and applied to all users through batch inference, using the same network parameters for every user. It comprises four fully connected hidden layers with $(256,128,128,64)$ neurons, each followed by batch normalization and ReLU activation. The five-neuron linear output layer estimates each user's three-dimensional position and two device orientation angles. A dataset of $N_{\mathrm{DNN}}=10^5$ network snapshots is generated, each containing $K=5$ users. 
The device orientation angles, $\varsigma_k$ and $\Theta_k$, are sampled according to their respective distributions within the considered angular ranges. Blockers are generated according to the model in Section~\ref{sec:sm}. The IRS configuration is kept fixed during dataset generation. The resulting dataset is divided by network snapshots into $80\%$ training, $10\%$ validation, and $10\%$ testing samples. Both input features and output targets are standardized to zero mean and unit variance. The DNN is trained using Adam with a learning rate of $\mathcal{E}_{\mathrm{DNN}}=10^{-3}$ and a mini-batch size of $64$ for $30$ epochs, after which the validation loss stabilizes. For the second stage, the DRL actor and critic networks each employ two fully connected hidden layers with $256$ and $128$ neurons, respectively. The actor uses a Tanh output layer to bound its continuous actions, which are subsequently mapped to their respective physical ranges. Both networks are optimized using Adam. The remaining learning parameters are listed in Table~\ref{tab:learning_params}.

\begin{figure}[t]
    \centering
    \includegraphics[width=0.45\textwidth]{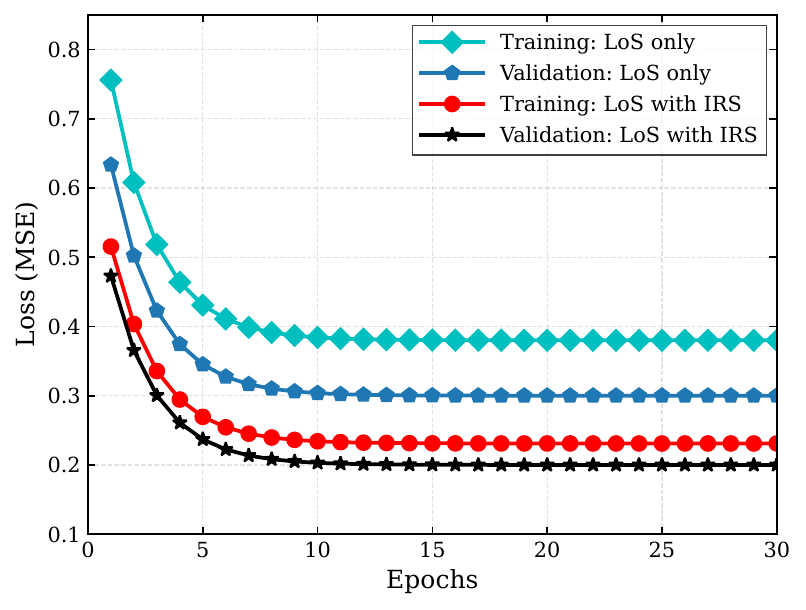}
    \caption{Training and validation losses of DNN.}
    \label{fig:Loss_mse}
    \vspace{-1mm}
\end{figure}

\vspace*{-5mm}
\subsection{Performance Analysis and Results}

First, we evaluate the DNN performance. Fig.~\ref{fig:Loss_mse} shows the training and validation losses in terms of the normalized MSE versus the epoch index, for the LoS-only and the LoS with IRS configurations, over the 30 training epochs used for training and validation. It can be seen that both losses decrease as the epoch index increases and stabilize within the first twelve epochs, which demonstrates that the obtained models are not overfitting and generalize well to unseen data. The validation loss is slightly lower than the training loss, reflecting the difference between the network's training and evaluation modes.  The LoS-only configuration achieves training and validation losses of approximately 0.38 and 0.30, respectively, whereas the LoS-with-IRS configuration achieves 0.23 and 0.20. Since both configurations use the same input dimension, architecture, and number of trainable parameters, this improvement is attributed to the additional information provided by the IRS-reflected paths. While the direct-path measurements may exhibit similar patterns for different user positions and orientations, the IRS reflections provide complementary spatial and angular information that improves the estimation performance.

\begin{figure}[t]
    \centering 
    \includegraphics[width=0.45\textwidth]{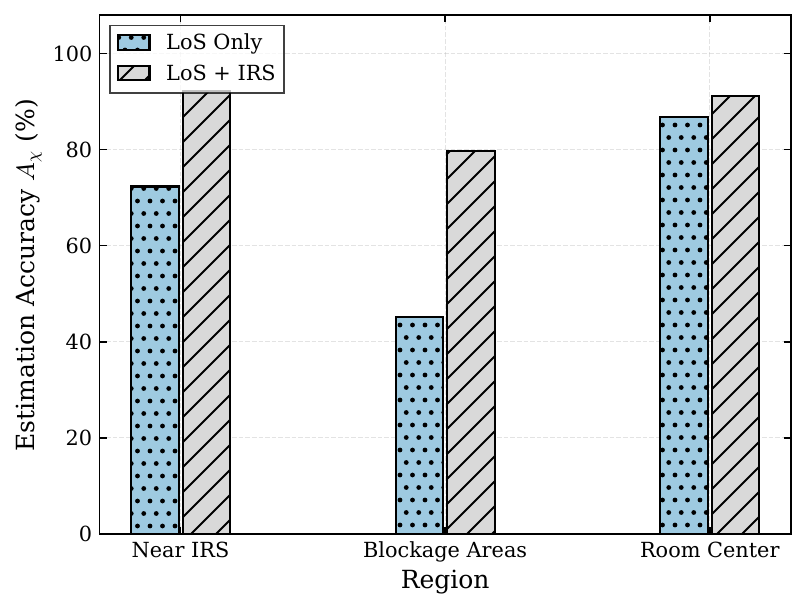}
    \caption{ Estimation accuracy in different regions.}
    \label{fig:accuracy}
    \vspace{-1mm}
\end{figure}

Fig.~\ref{fig:accuracy} shows the 3D position estimation accuracy at a fixed threshold of $\chi=20$~cm as defined in~\eqref{eq:accuracy}. The LoS with IRS configuration outperforms the LoS-only configuration in all three regions, with the improvement scaling with the geometric information added by the reflected NLoS signals. The accuracy increases from $72.3\%$ to $92.2\%$ in the near-IRS region, where the reflected paths provide additional measurement information. The largest improvement is observed in the blocked-link region, increasing from $45.2\%$ to $79.7\%$, demonstrating the benefit of IRS reflections when direct LoS paths are obstructed. In the room-center region, the improvement is smaller because the direct links from the surrounding APs provide relatively strong measurements.

\begin{figure}[!t]
    \centering
    \includegraphics[width=0.45\textwidth]{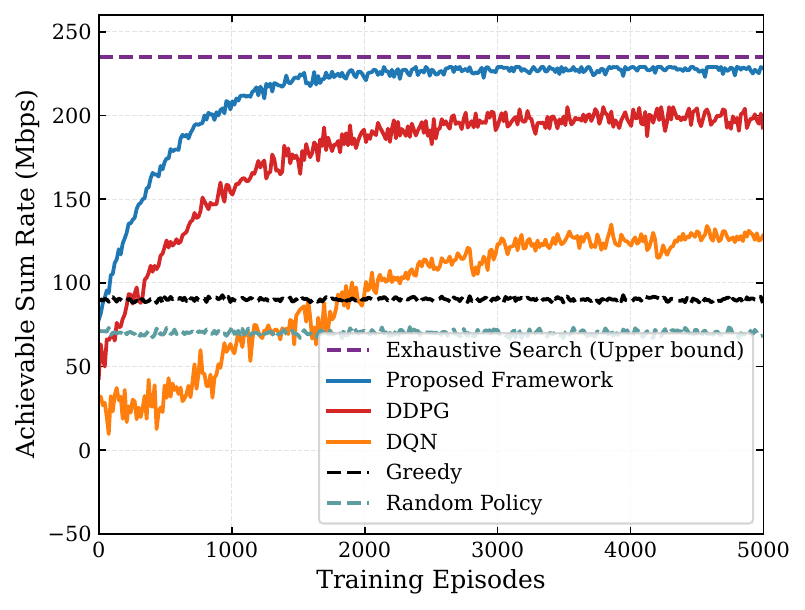}
    \caption{Sum rate per episode during the training for different algorithms.}
    \label{fig:Reward}
    \vspace{-1mm}
\end{figure}

The performance of the proposed framework is first evaluated in terms of the achievable sum rate over the training episodes, as shown in Fig.~\ref{fig:Reward}. The comparison includes two DRL algorithms, DDPG (Baseline 1, defined below) and DQN~\cite{Hammadi2022DQN}, the exhaustive-search benchmark, and the greedy~\cite{liu2023sum} and random policy as non-learning references. It can be seen that the proposed framework converges after approximately 2000 episodes to a stable sum rate of 225~Mbps, approximately 4.3\% below the exhaustive-search upper bound of 235~Mbps. Moreover, it outperforms DDPG and DQN by approximately 13\% and 79\%, respectively. This improvement is attributed to the DNN estimation stage, which provides the agent with user position and orientation estimates, allowing it to focus on optimizing power allocation and IRS orientations rather than learning the user state directly from raw measurements. This also accelerates convergence. In contrast, DQN converges more slowly and saturates at approximately 126~Mbps, with noticeable fluctuations, due to the limitations of its discretized action space in representing continuous power allocation and IRS orientation variables. Finally, the greedy and random policies achieve only $90$ and $70$~Mbps, respectively, as the greedy policy optimizes the immediate sum rate without considering future
decisions, while the random policy does not learn from previous interactions.

\setlength{\textfloatsep}{8pt}

Next, we evaluate the performance of the proposed framework against the following baselines.
\begin{itemize}[leftmargin=*,nosep]
\item \textbf{Baseline 1 (Standard DDPG):} A standard DDPG
algorithm~\cite{lillicrap2015continuous} solves \textbf{P1} without the first-stage estimation. The agent observes the raw per-AP measurement vector instead of the estimated positions and orientations, and jointly learns the user state and the resource allocation together. It optimizes the same power allocation and IRS mirror orientations as the proposed framework, with the induced decoding order. This comparison isolates the benefit of the DNN estimation stage.
\item \textbf{Baseline 2 (SCA):} Problem \textbf{P1} is solved using the SCA algorithm~\cite{tahira2019optimization}, with the same DNN-estimated user states. This baseline evaluates the performance of the proposed DRL policy against conventional iterative optimization, particularly under user mobility.

\item \textbf{Baseline 3 (No IRS):} The system is modeled without the IRS, and the proposed framework optimizes the NOMA power allocation over the LoS-only channel. The DNN therefore operates on the LoS-only feature vector; the IRS angles are removed from both the state and the action spaces, and the decoding order is determined by the LoS channel gains alone. This baseline evaluates the overall gain achieved by deploying the IRS. 

\item \textbf{Baseline 4 (OMA/TDMA):} The system operates under orthogonal multiple access using time-division multiple access (TDMA), where each user is served in a separate time slot of duration $T/K$. The full power budget $P_{\max}$ is used in each slot, and each user's achievable rate is therefore scaled by its time allocation of $1/K$. Since there is no power allocation or SIC to optimize, the proposed framework optimizes the IRS mirror orientations only using the same DNN estimates. This baseline evaluates the benefit of using NOMA over OMA. %
\end{itemize}
\begin{figure}[t]
    \centering
    \includegraphics[width=0.45\textwidth]{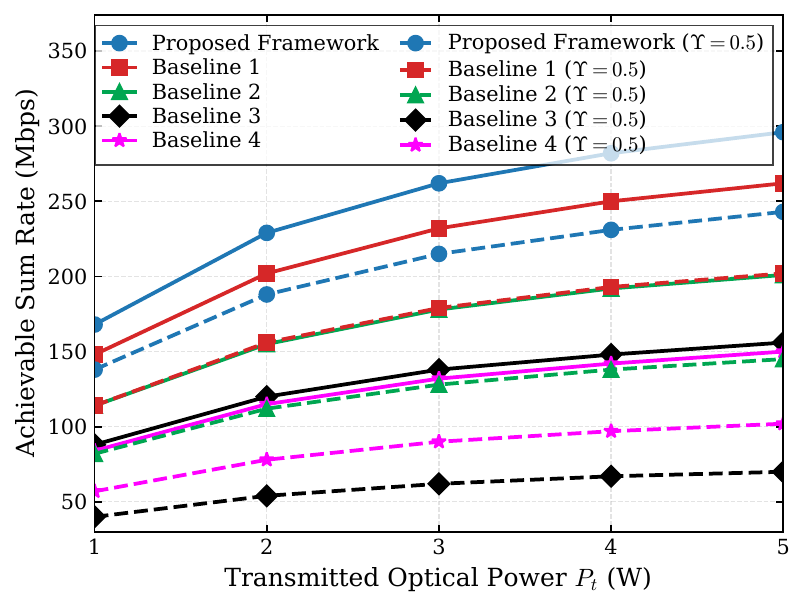}
\caption{ Sum rates versus the transmitted optical power under different blockage conditions. $M$ = 49.}
    \label{fig:blockage}
    \vspace{-1mm}
\end{figure}

Fig.~\ref{fig:blockage} shows the achievable sum rate against the transmitted optical power $P_{t}$ under blockage-free ($\Upsilon=0$, solid lines) and blocked ($\Upsilon=0.5$, dashed lines) conditions. It can be seen that the proposed framework achieves the highest sum rate across the considered power range, increasing from $168$~Mbps at $P_{t}=1$~W to $296$~Mbps at $P_{t}=5$~W. This improvement results from jointly optimizing the power and IRS orientations according to the estimated user state. Under blockage-free conditions, Baseline~1 remains approximately $12\%$ below the proposed framework, highlighting the benefit of the first-stage estimation. Baseline~2 achieved only $155$~Mbps at $P_{t}=2$~W, as the iterative SCA procedures may converge to a local solution. Moreover, without blockage and at $M=49$, Baseline~3 slightly outperforms Baseline~4, since NOMA serves users over the full transmission bandwidth, while TDMA allocates only a fraction of the frame to each user. The benefit of the IRS becomes more evident under blockage. At $\Upsilon=0.5$ and $P_t=2$~W, the proposed framework experiences a sum-rate reduction of only, while Baseline~3 drops by $55\%$ due to the absence of reflected paths. Consequently, Baseline~4 outperforms Baseline~3 under blockage, with a smaller rate reduction of $32\%$. This demonstrates that combining IRS-assisted transmission with orthogonal access can preserve a higher sum rate than relying on NOMA over direct links alone when LoS paths are obstructed. Finally, Baselines~1 and~2 experience reductions of approximately $23\%$ and $28\%$, respectively, further demonstrating the robustness of the proposed framework under blockage.

\begin{figure}[t]
    \centering
    \includegraphics[width=0.45\textwidth]{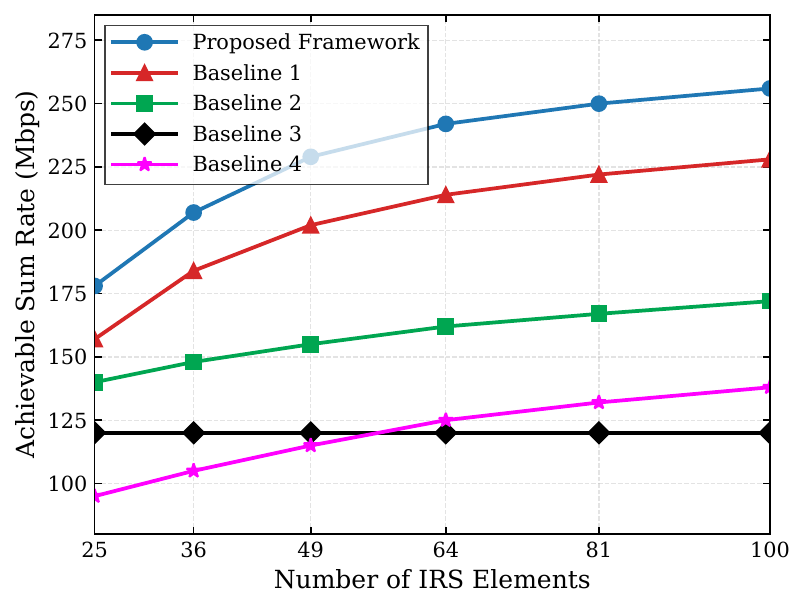}
    \caption{Sum rates versus the number of IRS reflective mirrors $M$. Transmitted optical power, $P_t= 2 W$.}
    \label{fig:sum_rate_vs_irs}
    \vspace{-1mm}
\end{figure}

Fig.~\ref{fig:sum_rate_vs_irs} shows the achievable sum rate against the number of IRS mirrors, ranging from $25$ to $100$. It can be seen that the sum rate of all IRS-aided schemes increases with the number of mirrors, as a larger reflecting aperture provides more controllable elements and greater flexibility in shaping the effective channel gains. The proposed framework achieves the highest sum rate over the whole range, reaching $256$~Mbps at $100$ elements, which is approximately $12\%$ above Baseline~1 and $49\%$ above Baseline~2. In contrast, Baseline~3 remains constant at $120$~Mbps, as it does not employ an IRS and therefore cannot benefit from the additional elements. Moreover, Baseline~4 overtakes Baseline~3 when the number of mirrors increases beyond $49$, indicating that an optimized IRS can compensate for the limitations of the direct-link-only configuration, even under orthogonal access.

\begin{figure}[t]
    \centering
    \includegraphics[width=0.45\textwidth]{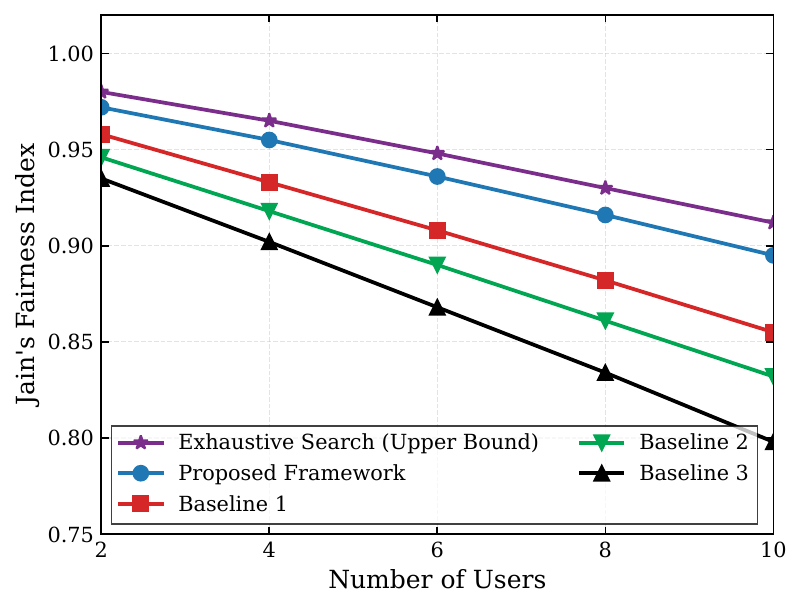}
    \caption{ Fairness index versus different numbers of users.}
    \label{fig:fairness}
    \vspace{-1mm}
\end{figure}

Fairness is another important performance metric in NOMA systems, where power allocation depends on the users' channel conditions~\cite{Hammadi2022DQN}. Fig.~\ref{fig:fairness} shows Jain's fairness index, computed over the users' achievable rates, against the number of users. The comparison focuses on the NOMA-based schemes. It can be seen that the fairness of all schemes decreases as the number of users increases, since more users compete for the available resources. The proposed framework closely follows the upper bound, achieving $0.895$ at $K=10$, which corresponds to approximately $98.1\%$ of the upper-bound value. This is attributed to its QoS-weighted objective, which promotes a more balanced power allocation among users. Baseline~1 achieves $0.855$, highlighting the benefit of the first-stage estimation, while Baseline~2 achieves $0.833$, as the SCA-based optimization provides a less balanced solution. Finally, Baseline~3 performs worst, reaching $0.798$, approximately $10.8\%$ below the proposed framework, since users experiencing blocked direct links cannot benefit from IRS-reflected paths.

\begin{figure}[t]
    \centering
    \includegraphics[width=0.45\textwidth]{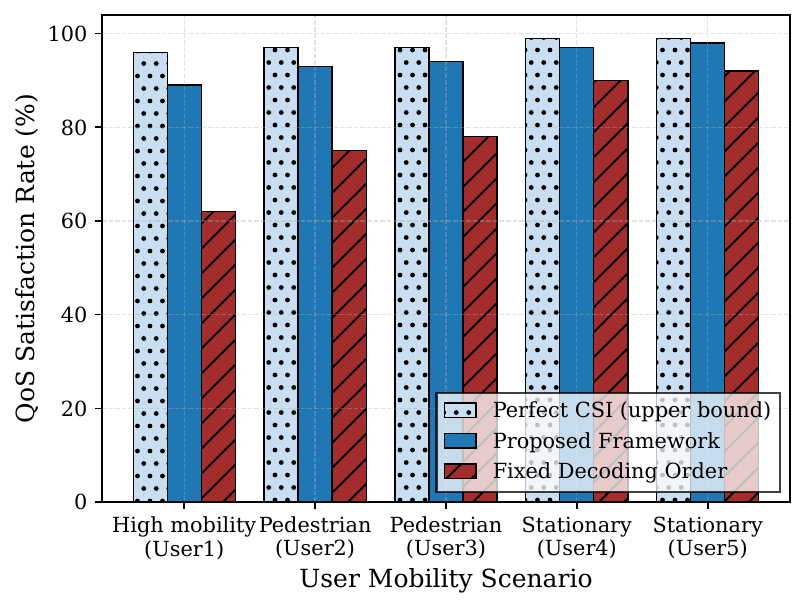}
    \caption{QoS satisfaction for different user mobility scenarios.}
    \label{fig:qos_fair_comparison}
    \vspace{-1mm}
\end{figure}

Following recent works on QoS satisfaction in dynamic VLC-NOMA networks~\cite{alhameed2025marl}, we evaluate the ability of the proposed framework to satisfy users' rate requirements under different mobility conditions. The QoS satisfaction rate of user $k$ is defined as the average fraction of time its achievable rate meets the minimum required rate. Fig.~\ref{fig:qos_fair_comparison} compares the proposed framework with a fixed decoding-order scheme and a perfect-CSI upper bound under the same QoS requirements. In the proposed framework, the decoding order is updated at every time step according to the effective channel gains after IRS reconfiguration, whereas the fixed-order scheme determines the decoding order from the channel gains at the beginning of each episode. The stationary users achieve satisfaction rates above $90\%$ under all three schemes, despite having the highest rate requirements, as their channel conditions remain relatively stable. As user mobility increases, the proposed framework maintains performance within $7\%$ points of the perfect-CSI upper bound. In contrast, the fixed-order scheme experiences a substantial reduction, with satisfaction decreasing from $90\%$ for stationary users to $62\%$ for the high-mobility user. This is because user mobility changes the effective channel-gain ranking, making the initially determined decoding order increasingly unsuitable. By updating the decoding order at every time step, the proposed framework adapts to these changes and maintains higher QoS satisfaction.
\vspace*{-1mm}
\section{Conclusions} \label{Sec:6.7}
In this paper, the joint optimization of power allocation and IRS mirror orientation was formulated to maximize the sum rate while satisfying the QoS requirements of multiple users in a dynamic indoor IRS-assisted NOMA-OWC system. The key novelty lies in leveraging the IRS to shape the effective channel gains, from which the decoding order is induced and updated according to the users' channel conditions rather than being fixed in advance. To solve this problem, a novel two-stage DNN-DRL framework was proposed. In the first stage, a DNN estimates the users' positions and orientations from the received signal strengths over both the LoS and IRS-reflected links. In the second stage, a DPG-based DRL algorithm uses these estimates to jointly optimize the power allocation and IRS mirror orientations, with the decoding order induced from the resulting effective channel gains. Numerical results demonstrated that incorporating the IRS-reflected NLoS components significantly improves position estimation, particularly under blockage, where the accuracy increases from $45.2\%$ to $79.7\%$. Furthermore, updating the decoding order at every time step maintains QoS satisfaction close to the perfect-CSI benchmark across different mobility scenarios, whereas a fixed decoding order suffers considerable degradation under high mobility. The proposed framework also approaches the upper bounds in both sum rate and fairness, achieving a sum rate within approximately $4.3\%$ of the exhaustive-search upper bound and $98.1\%$ of the fairness upper bound. Overall, it outperforms conventional convex optimization, standard DRL baselines, and configurations without the IRS or using orthogonal access.

\vspace{-2mm}

\begingroup
\footnotesize
\setlength{\itemsep}{0pt}
\setlength{\parskip}{0pt}
\setlength{\parsep}{0pt}
\renewcommand{\baselinestretch}{0.9}
\bibliographystyle{IEEEtran}
\bibliography{References}
\endgroup
\end{document}